\documentclass[aps,preprint,a4paper,showpacs,showkeys,superscriptaddress]{revtex4-1}

\usepackage{latexsym}
\usepackage{amsmath,amssymb}
\usepackage{graphicx}
\usepackage{subfigure}
\usepackage{cancel}
\usepackage{hyperref}     
\hypersetup{colorlinks,%
  citecolor=blue,%
  linkcolor=cyan,%
  pdftex}

\begin{document}


\title{The role of the cosmological constant as a pressure in the (2+1)-dimensional black string}

\author{Yongwan Gim}%
\email[]{yongwan89@sogang.ac.kr}%
\affiliation{Department of Physics, Sogang University, Seoul, 04107,
  Republic of Korea}%
\affiliation{Research Institute for Basic Science, Sogang University,
  Seoul, 04107, Republic of Korea} %

\author{Wontae Kim}%
\email[]{wtkim@sogang.ac.kr}%
\affiliation{Department of Physics, Sogang University, Seoul, 04107,
  Republic of Korea}%

\date{\today}

\begin{abstract}
It has been claimed that
the cosmological constant in AdS black holes such as the BTZ black hole
plays the role of the thermodynamic variable of a pressure
in the thermodynamic first law and the Smarr relation from the scaling law
of the Christodoulou-Ruffini formula.
However, the dual solution of the BTZ black hole is the black string
which is asymptotically flat despite the
presence of the cosmological constant, and so
the explicit form of the pressure with the role of the cosmological constant is
unclear in the black string
since the pressure is subject to the choice of the energy-momentum tensor.
Thus,
we show that if the pressure of the black string is still assumed to
be proportional to the cosmological constant similar to the case of the BTZ black hole,
then the thermodynamic first law is consistent with the Smarr relation
from the Christodoulou-Ruffini formula,
and the thermodynamic quantities for the pressure
are well-behaved under the dual transformation.
\end{abstract}



\maketitle


\section{Introduction}
\label{sec:intro}

It has been claimed that the cosmological constant in
anti-de Sitter (AdS) black holes plays a role of a thermodynamic variable \cite{Henneaux:1984ji, Teitelboim:1985dp},
which is consequently interpreted as a bulk pressure \cite{Creighton:1995au, Caldarelli:1999xj,  Kastor:2009wy, Frassino:2015oca}
and its conjugate variable is also identified with
the thermodynamic volume \cite{Dolan:2010ha, Cvetic:2010jb}.
If the cosmological constant were not regarded as the pressure
in the framework of the scaling law from the Christodoulou-Ruffini formula,
then the Smarr relation \cite{Smarr:1972kt} could not be defined in the asymptotically AdS black holes
although the thermodynamic first law could be well-defined without the pressure \cite{Gibbons:2004ai}.
In this respect, it appears to be plausible to regard the cosmological constant as a thermodynamic variable of
the pressure in order for the uniform and consistent formulation of the Smarr relation along
with the thermodynamic first law \cite{Kubiznak:2016qmn}.
Expectedly, the cosmological constant in
the Ba${\rm \tilde{n}}$ados-Teitelboim-Zanelli (BTZ) black hole \cite{Banados:1992wn}
plays a definite role of the pressure in the Smarr relation and the thermodynamic first law \cite{Frassino:2015oca}.
On the other hand, the dual description of the BTZ black hole in the framework of the low energy string theory \cite{Horowitz:1993jc, Horowitz:1993wt} provides the three-dimensional black string; however,
it is asymptotically flat
in spite of the presence of the cosmological constant in contrast to the BTZ black hole.
Thus, one might wonder what the thermodynamic role of the cosmological constant in the black string is.


In this work,
we  find
the pressure of the black string from the same energy-momentum tensor as that of the BTZ black hole; however,
the conjugate variables to the mass, the charge, and the pressure
take non-standard forms.
Instead, assuming that the energy-momentum tensor of the black string
comes from only the term of the cosmological constant,
we can identify the suitable form of the pressure compatible with
the standard definitions of the Hawking temperature, the potential, and the volume.
Eventually, the pressure of the black string becomes to be
proportional to that of the BTZ black hole.
Employing this pressure, we derive the thermodynamic first law in the Padmanabhan's formalism \cite{Padmanabhan:2002sha, Padmanabhan:2012gx}
and the Smarr relation from the Christodoulou-Ruffini formula \cite{Christodoulou:1972kt}.
We also show that the advantage of the latter choice for the pressure is that
thermodynamic quantities are well-behaved under the dual transformation.

 In Sec.~\ref{sec:dual1},
we recapitulate the duality 
in the three-dimensional low energy string theory presented in Refs.~\cite{Horowitz:1993jc, Horowitz:1993wt}.
In Sec.~\ref{sec:padBTZ}, we revisit the thermodynamics for the BTZ black hole
at the inner and outer horizons in order to identify the corresponding pressures by using the Padmanabhan's method.
In Sec.~\ref{sec:padBS},
we choose
the suitable form of the pressure of the black string
by means of the Padmanabhan's method, and derive the thermodynamic first law from the equations of motion.
Next, the Smarr relation is obtained from the scale invariance of the Christodoulou-Ruffini formula  by employing the pressure.
In Sec.~\ref{sec:dual}, we study the dual relation of the thermodynamic quantities
between the BTZ black hole and the black string.
Finally, conclusion and discussion will be given in Sec.~\ref{sec:sum}.

\section{Dual solutions}
\label{sec:dual1}
Let us start with the action given as \cite{Horowitz:1993jc, Horowitz:1993wt},
\begin{equation}\label{eq:action}
{\cal I}=\frac{1}{2\pi}\int d^3x\sqrt{-g}e^{-2\phi}\left[ \frac{4}{\ell^2} +R+4(\nabla \phi)^2-\frac{1}{12}H_{\mu\nu\rho}H^{\mu\nu\rho}\right],
\end{equation}
where $\phi$ is the dilaton field, $H_{\mu\nu\rho}$ is the three-form field strength
of $H_{\mu\nu\rho}=\partial_\mu B_{\nu \rho}+\partial_\nu B_{\rho \mu}+\partial_\rho B_{\mu \nu}$, and
the cosmological constant is $\Lambda =-{\ell^{-2}}$.
The equations of motion for the action \eqref{eq:action} are given by
\begin{align}
&G_{\mu\nu}=R_{\mu\nu}-\frac{1}{2}g_{\mu\nu} R=8\pi G T_{\mu\nu}, \label{eq:EOMpad1}\\
&\nabla^\mu(e^{-2\phi} H_{\mu\lambda\rho} )=0,  \qquad R-\frac{1}{12}H^2+4\nabla^2 \phi - 4(\nabla \phi)^2 +\frac{4}{\ell^2}=0, \label{eq:EOM3} \\
& T_{\mu\nu}=\frac{1}{8\pi G}\left[-2\nabla_\mu \nabla_\nu \phi +\frac{1}{4}H_{\mu\lambda\rho}H_{\nu}^{\lambda\rho} +g_{\mu\nu}\left(\frac{2}{\ell^2}+2\nabla^2 \phi -2(\nabla \phi)^2 -\frac{1}{24} H^2 \right)\right] \label{eq:EOMpad2}
\end{align}
with respect to each fields, and $8G = 1$ for convenience.
One of the solutions to Eqs.~\eqref{eq:EOMpad1} and \eqref{eq:EOM3}  is the BTZ black hole described by
\begin{align}
ds^2 &=-\left(\frac{r^2}{\ell^2} -M\right)dt^2 -J dt d\varphi+ \frac{1}{\frac{r^2}{\ell^2}-M+\frac{J^2}{4r^2}}dr^2+r^2d\varphi^2, \label{eq:BTZ1}\\
B_{\varphi t} &=r^2 /\ell^2,\quad \phi=0,  \label{eq:BTZ2}
\end{align}
where the inner and outer horizons are $r_\pm^2 = (\ell/2)\left(M\ell \pm \sqrt{M^2\ell^2-J^2}\right)$.
If we consider $B=0$ as well as $\phi=0$,
the Lagrangian \eqref{eq:action} becomes ${\cal L}=R+4\Lambda$
which is different from the conventional Lagrangian with the cosmological constant of ${\cal L}=R+2\Lambda$.
The mass and the angular momentum for the BTZ black hole are expressed as
$M = (r_{+}^2 + r_{-}^2)/\ell^2$ and $\quad J=2r_{+}r_{-}/\ell$.
The angular velocity $\Omega_+$, the Hawking temperature $T_+$, and the entropy $S_+$ on the outer horizon are also given as
\begin{align}
 \Omega_+ = \frac{J}{2r_+^2}, \quad
  T_+=\frac{r_+^2-r_-^2}{2\pi \ell^2 r_+}, \quad S_+ = 4\pi r_+. \label{eq:TnSBTZ}
\end{align}

Next, performing the dual transformation of Eqs. \eqref{eq:BTZ1} and \eqref{eq:BTZ2}
\cite{Buscher:1987sk, Buscher:1987qj} as
 \begin{align}
&\tilde{g}_{\varphi\varphi} =\frac{1}{g_{\varphi \varphi}},\qquad
\tilde{g}_{\varphi \mu}=\frac{B_{\varphi \mu}}{g_{\varphi\varphi}}, \qquad
\tilde{g}_{\mu\nu}=g_{\mu \nu} - (g_{\varphi \mu} g_{\varphi \nu}-B_{\varphi \mu}B_{\varphi \nu})/g_{\varphi\varphi},\\
&\tilde{B}_{\varphi \mu} =\frac{g_{\varphi \mu}}{g_{\varphi \varphi}},\qquad
\tilde{B}_{\mu\nu}=B_{\mu\nu}-\frac{2}{g_{\varphi\varphi}}g_{\varphi [\mu}g_{\nu]\varphi},\qquad
\tilde{\phi}=\phi-\frac{1}{2}\ln g_{\varphi\varphi},
 \end{align}
where $\mu,~\nu$ run over all coordinates except $\varphi$,
one can obtain
the charged black string solution after some diagonalization,
\begin{align}
d\tilde{s}^2 &=-\left(1-\frac{\cal M}{\hat{r}}\right)d\hat{t}^2+ \frac{\ell^2}{4\hat{r}^2} \frac{1}{\Big(1-\frac{\cal M}{\hat{r}}\Big) \left(1-\frac{{\cal Q}^2}{{\cal M}\hat{r}}\right)}d\hat{r}^2+\left(1-\frac{{\cal Q}^2}{{\cal M}\hat{r}}\right)d\hat{x}^2, \label{eq:BS}   \\
\tilde{B}_{\hat{x}\hat{t}} &=\frac{{\cal Q}}{\hat{r}},\quad  \tilde{\phi}=-\frac{1}{2}\ln(\hat{r}\ell), \notag
\end{align}
where
the outer horizon is $\hat{r}_+ = {\cal M}$ and the inner horizon is $\hat{r}_- = {\cal Q}^2/{\cal M}$.
 The geometry \eqref{eq:BS} is asymptotically flat
since the square of the Riemann curvature  vanishes at the asymptotic infinity.
 The mass ${\cal M}$ and the charge $ Q$ per unit length of the black string
are related to the outer horizon and the angular momentum of the BTZ black hole
as ${\cal M}=r_+^2/\ell$ and ${\cal Q}=J/2$.
The potential $\Phi$ of the charge of the black string can be read off from the field $\tilde{B}_{\hat{x}\hat{t}}$ at the outer horizon as
\begin{equation}\label{eq:Phi}
\Phi_+ =\tilde{B}_{\hat{x}\hat{t}}|_{\hat{r}=\hat{r_+}} = \frac{ {\cal Q}}{\hat{r}_+}.
\end{equation}
The Hawking temperature and entropy are given as \cite{Horowitz:1993wt}
\begin{equation}\label{eq:TBS}
{\cal T}_+ = \frac{\sqrt{{\cal M}^2 - {\cal Q}^2}}{2\pi \ell {\cal M}},~~~~~
{\cal S}_{\rm bs}^+
=2 \ell \sqrt{{\cal M}^2- {\cal Q}^2} \hat{X},
\end{equation}
where $\hat{X}=\int d\hat{x}$ is the comoving length of the black string \cite{Kaloper:1998vw}.
Here, the entropy per unit length is simply written as
$s_+={\cal S}_{\rm bs}^+ /\hat{X}
 =2 \ell \sqrt{{\cal M}^2- {\cal Q}^2}$.

\section{Thermodynamic first law and Smarr relation in the BTZ black hole}
\label{sec:padBTZ}

The cosmological constant $\Lambda$ is identified with the bulk pressure
 and its conjugate variable can be treated as the thermodynamic volume
 \cite{Creighton:1995au, Caldarelli:1999xj,  Kastor:2009wy, Dolan:2010ha, Cvetic:2010jb, Frassino:2015oca, Detournay:2012ug, Liang:2017kht, Bhattacharya:2017hfj, Cvetic:2018dqf}.
Then, the cosmological constant gives rise to the pressure term of $VdP$
in the thermodynamic first law at the outer horizon in the BTZ black hole
\cite{Kastor:2009wy, Frassino:2015oca}.
In order to identify the pressure at the outer and inner horizons,
we consider the Padmanabhan's method to get the thermodynamic first laws \cite{Padmanabhan:2002sha, Padmanabhan:2012gx},
and obtain the consistent Smarr relations with the thermodynamic
first laws.

We rewrite the Hawking temperature of the BTZ black hole in Eqs.~\eqref{eq:TnSBTZ}  as
\begin{equation}\label{eq:Tout_BTZ}
T_+ = \frac{h'(r_+)}{4\pi}-\frac{r_-^2}{2\pi \ell^2 r_+},
\end{equation}
where $h(r)=-g_{tt}=\ell^{-2}(r^2-r_+^2-r_-^2)$.
From the field equation of $G^r_r=\pi T^r_r$ in Eq. \eqref{eq:EOMpad1},
one can obtain
\begin{align}\label{eq:pad11}
-\frac{2r_-^2}{\ell^2 r_+}dr_+  &=\left(\frac{h'(r_+)}{4\pi}-\frac{r_-^2}{2\pi \ell^2 r_+}\right)d(4\pi r_+)- T^r_r d(\pi r_+^2)
\end{align}
at the outer horizon after multiplying $dr_+$ by using the Padmanabhan's method  \cite{Padmanabhan:2002sha, Padmanabhan:2012gx}.
Then,  the standard first law is expressed as
\begin{equation}\label{eq:pad111}
dE_+= T_+dS_+ + \Omega_+ dJ -PdV_+,
\end{equation}
 where the energy $E_+$ is defined as $E_+=r_-^2/\ell^2 $.
Here, the pressure of the BTZ black hole is naturally derived from the energy momentum tensor as
\cite{Creighton:1995au, Caldarelli:1999xj,  Kastor:2009wy}
\begin{equation}\label{eq:P_BTZ}
T^r_r=P=\frac{1}{\pi \ell^2},
\end{equation}
 and its conjugate variable of
$V_+ = \pi r_+^2$ is
the thermodynamic volume \cite{Dolan:2010ha, Cvetic:2010jb}.

Now, separating the left hand side of Eq.~\eqref{eq:pad11} into the mass part and the angular part,
one can obtain
\begin{align}
d\left(\frac{r_+^2+r_-^2}{\ell^2}\right)=\left(\frac{h'(r_+)}{4\pi}-\frac{r_-^2}{2\pi \ell^2 r_+}\right)d(4\pi r_+) +\frac{r_-}{\ell r_+}d\left(\frac{2r_-r_+}{\ell}\right)+ \pi r_+^2 dP, \label{eq:pad12}
\end{align}
and consequently it becomes
\begin{align}
dM =T_+ dS_++\Omega_+ dJ+V_+dP. \label{eq:first_BTZ}
\end{align}
Note that
the Padmanbhan's approach and the extended phase space approach are quite different in interpretations as shown in Refs.~\cite{Hansen:2016ayo, Kubiznak:2016qmn}.
In the former case, the energy  is the internal energy of the thermal system and the pressure is just an intensive thermodynamic quantity,
while in the latter case
the energy  is the enthalpy of the thermal system and the pressure is treated as a thermodynamic variable;
it turns out that the enthalpy  is nothing but the ADM mass.
As a matter of fact,
Eq.~\eqref{eq:pad111} and Eq.~\eqref{eq:first_BTZ} are related to each other
through  the transformation  of $M=E_++PV_+$,
which corresponds to the Legendre projection in Refs.~\cite{Hansen:2016wdg, Hansen:2016gud}.

On the other hand,
black holes may have an inner horizon as well as the outer horizon.
In fact, the quantized charges of various black holes are
related to the product of the entropies of the inner and outer horizons \cite{Larsen:1997ge, Cvetic:2010mn,  Detournay:2012ug, Cvetic:2018dqf}, which gives a looking glass for probing the microscopic properties of black holes \cite{Cvetic:2010mn}.
For the inner mechanics of the BTZ black hole,
it was shown that the thermodynamic first law is satisfied
without the pressure term~\cite{ Detournay:2012ug} and
the negative temperatures also arise inevitably on inner horizons~\cite{Cvetic:2018dqf}.
Now, let us derive the thermodynamic first law on the inner horizon by taking into account the pressure term.

The temperature and the entropy on the inner horizon of the BTZ black hole are given as
\begin{align}
T_- = \frac{h'(r_-)}{4\pi}-\frac{r_+^2}{2\pi \ell^2 r_-}=\frac{r_-^2-r_+^2}{2\pi \ell^2 r_-}, \qquad S_- =4\pi r_-, \label{eq:Sin_BTZ}
\end{align}
where the entropy was obtained by the area law.
In the Padmanabhan's method \cite{Padmanabhan:2002sha, Padmanabhan:2012gx},
the equation of motion of $G^r_r=\pi T^r_r$ at the inner horizon
is written as
\begin{align}\label{eq:1st_Pad_inner}
-\frac{2r_+^2}{\ell^2 r_-}dr_-  &=\left(\frac{h'(r_-)}{4\pi}-\frac{r_+^2}{2\pi \ell^2 r_-}\right)d(4\pi r_-)- T^r_r d(\pi r_-^2),
\end{align}
Then,  the standard first law is expressed as
\begin{equation}\label{eq:}
dE_-= T_-dS_- +\Omega_- dJ -PdV_- ,
\end{equation}
 where the energy $E_-$ is defined as $E_-=r_+^2/\ell^2 $.

Next, by considering the pressure as a thermodynamic variable, we can rewrite Eq.~\eqref{eq:1st_Pad_inner} as
\begin{align}
d\left(\frac{r_+^2+r_-^2}{\ell^2}\right)&=\left(\frac{h'(r_-)}{4\pi}-\frac{r_+^2}{2\pi \ell^2 r_-}\right)d(4\pi r_-)+\frac{r_+}{\ell r_-}d\left(\frac{2r_+r_-}{\ell}\right) +(\pi r_-^2 )dT^r_r,\notag
\end{align}
and then one can obtain the thermodynamic first law at the inner horizon as
\begin{equation}\label{eq:first_in_BTZ}
dM=T_- dS_- +\Omega_- dJ+V_- dP,
\end{equation}
where the angular velocity $\Omega_-$ and the conjugate variable $V_-$ to the pressure at the inner horizon are defined as $\Omega_- = r_+/\ell r_- $ and $V_- = \pi r_-^2$.
We regarded the cosmological constant as the pressure,
and  admitted the negative temperature at the inner horizon.
Note that
Eq. \eqref{eq:first_in_BTZ} reduces to
$dM=-T^{\rm D}_- dS_- +\Omega_- dJ$ in Ref.~\cite{Detournay:2012ug}
if $dP=0$ and $T_- = -T^{\rm D}_-$.

Now, let us study the Smarr relations of the BTZ black hole on the outer and inner horizons compatible with the
thermodynamic first laws
\eqref{eq:first_BTZ} and \eqref{eq:first_in_BTZ}\cite{Kastor:2009wy, Frassino:2015oca, Kubiznak:2016qmn, Liang:2017kht}.
From the Christodoulou-Ruffini formula for the BTZ black hole given as \cite{Frassino:2015oca}
\begin{equation}\label{eq:CRBTZ}
M=\frac{{S_{\pm}}^2P}{16\pi}+\frac{4\pi^2 J^2}{{S_{\pm}}^2},
\end{equation}
the conjugate variables of $(T_\pm,\Omega,V_\pm)$ to the thermodynamic quantities $(S_\pm,J,P)$ are defined by
\begin{align}
T_\pm &= \left. \frac{\partial M}{\partial S_\pm} \right|_{J,P} = \frac{P S_\pm}{8\pi}-\frac{8\pi^2 J^2}{S_\pm^3}, \notag \\
\Omega_\pm &= \left. \frac{\partial M}{\partial J} \right|_{S_\pm,P}=\frac{8\pi^2 J}{S_\pm^2}, \label{eq:intensive_BTZ} \\
V_\pm &= \left. \frac{\partial M}{\partial P} \right|_{S_\pm,J} = \frac{S_\pm^2}{16\pi}. \notag
\end{align}
Note that $T_-$ in Eqs.~\eqref{eq:intensive_BTZ} is the same as the temperature in Eqs.~\eqref{eq:Sin_BTZ} and
it is negative~\cite{Cvetic:2018dqf}.
Taking the mass function \eqref{eq:CRBTZ} to be a homogeneous function of $M=M(S_\pm,J,P)$ under the scale transformation of $(M,S_\pm,J, P) \rightarrow ( M, \lambda S_\pm, \lambda J,\lambda^{-2} P)$,
we get the Smarr relation as
\begin{equation}\label{eq:Smarr_BTZ}
0= T_\pm S_\pm +\Omega_\pm J-2V_\pm P,
\end{equation}
which is compatible with the thermodynamic first laws
in the sense that the differentiation of Eq. \eqref{eq:Smarr_BTZ} with Eq. \eqref{eq:CRBTZ} gives Eqs. \eqref{eq:first_BTZ} and \eqref{eq:first_in_BTZ}.


\section{Thermodynamic first law and Smarr relation in black string}
\label{sec:padBS}

The black string of being
dual to the BTZ black hole also has the cosmological constant,
but it is asymptotically flat.
Thus, one might wonder how
the cosmological constant of the black string plays the role of the pressure
in the thermodynamic first law and the Smarr relation in the black string.
Let us study the thermodynamic first law by using the Padmanabhan's method
and the Smarr relation from the scaling property of the Christodoulou-Ruffini formula.

\subsection{Thermodynamic first law from Padmanabhan's method}

First of all,
let us obtain the thermodynamic quantities of the black string by using the dual transformation.
The coordinate $\hat{x}$ along the black string  \eqref{eq:BS} should  be periodic
since the angular coordinate $\varphi$ of the BTZ black hole \eqref{eq:BTZ1} is periodic.
In this regard, the whole string can be divided into an infinite array of the identical periodic string,
the so-called elementary black string, which is dual to the BTZ black hole \cite{Kaloper:1998vw}.
Thus, the coordinate $\hat{x}$ should be compactified on a circle of circumference of $\hat{X}$
calculated as
\begin{equation}\label{eq:X_BS}
\hat{X}=2\pi \sqrt{\frac{{\cal M}}{\ell({\cal M}^2 - Q^2)}}  = \frac{4\pi {\cal M}_{\rm tot}}{\ell \left({\cal M}_{\rm tot}^2-{\cal Q}_{\rm tot}^2\right)},
\end{equation}
where  the total mass and charge of the black string
are defined by the mass ${\cal M}$ and the axion charge ${\cal Q}$ per unit length as
${\cal M}_{\rm tot} = \alpha {\cal M} \hat{X},~ {\cal Q}_{\rm tot} = \alpha {\cal Q} \hat{X}$
with a calibration factor of $\alpha=1/\pi$.
By plugging Eq.~\eqref{eq:X_BS} into the entropy in Eq.~\eqref{eq:TBS},
the total entropy of the black string is obtained as
\begin{align}\label{eq:entropy_X}
  {{\cal S}_{\rm bs}^+}=2\pi \ell \sqrt{{\cal M}_{\rm tot}^2-{\cal Q}_{\rm tot}^2},
\end{align}
which is the same as the entropy of the BTZ black hole in Eqs.~\eqref{eq:TnSBTZ}  \cite{Horowitz:1993wt, Ho:1997uk}.
Further, we can also rewrite the temperature \eqref{eq:TBS} as
\begin{align}
{\cal T}_+ =\frac{\hat{r}_+ \sqrt{{\cal M}_{\rm tot}^2-{\cal Q}_{\rm tot}^2}f'(\hat{r}_+)}{2 \ell \hat{X}} =\frac{\sqrt{{\cal M}_{\rm tot}^2-{\cal Q}_{\rm tot}^2}}{2\pi \ell {\cal M}_{\rm tot}}, \label{eq:}
\end{align}
where $f(r)=1-{\cal M}/\hat{r}$.

Now, let us consider the equation of motion \eqref{eq:EOMpad1}
with the energy-momentum tensor \eqref{eq:EOMpad2}.
The equation of motion $G^{\hat{r}}_{\hat{r}}=\pi T^{\hat{r}}_{\hat{r}}$  multiplied by $d\hat{r}_+$
gives
\begin{equation}\label{eq:Pad1_BS}
\frac{\hat{r}_+^2 {\cal Q}_{\rm tot}^2 \hat{X}}{\pi \ell^2 {\cal M}_{\rm tot}^3}f'(\hat{r}_+) d\hat{r}_+ = \pi T^{\hat{r}}_{\hat{r}} d\hat{r}_+
\end{equation}
at the outer horizon,
where the pressure of the black string is found as
\begin{equation}\label{eq:P_BS}
T^{\hat{r}}_{\hat{r}}={\cal P}=\frac{{\cal Q}_{\rm tot}^2}{\pi \ell^2 {\cal M}_{\rm tot}^2}.
\end{equation}
Note that the pressure \eqref{eq:P_BS} relies on the total mass and charge as well as the cosmological constant.
Using this pressure,
 we can express Eq.~\eqref{eq:Pad1_BS} in terms of the analogous form of the thermodynamic first law as
\begin{align}
d{\cal M}_{\rm tot} &= \frac{{\cal M}_{\rm tot}\sqrt{{\cal M}_{\rm tot}^2-{\cal Q}_{\rm tot}^2}}{2\pi \ell {\cal Q}_{\rm tot}^2} d{{\cal S}_{\rm bs}^+} + \frac{{\cal M}_{\rm tot}(2{\cal Q}_{\rm tot}^2-{\cal M}_{\rm tot}^2)}{{\cal Q}_{\rm tot}^3}d{\cal Q}_{\rm tot}
+\frac{\pi \ell^2 {\cal M}_{\rm tot}^3 ({\cal M}_{\rm tot}^2-{\cal Q}_{\rm tot}^2)}{2{\cal Q}_{\rm tot}^4}d{\cal P}    \nonumber \\
&= \tilde{{\cal T}}_+ d{{\cal S}_{\rm bs}^+}+\tilde{\Phi}_+ d{\cal Q}_{\rm tot} + \tilde{{\cal V}}_+d{\cal P}, \label{eq:Pad3_BS}
\end{align}
where $\tilde{{\cal V}}_+$ is the conjugate variable to the pressure ${\cal P}$.
However,
the potential $\tilde{\Phi}_+$ and the temperature $\tilde{{\cal T}}_+$ in Eqs.~\eqref{eq:Pad3_BS} are different from the potential \eqref{eq:Phi} of the axion charge defined by the two-form field
and the Hawking temperature \eqref{eq:TBS} defined by the surface gravity.
Actually, there are no preferred forms of the pressure for the black string yet, unlike the case of the BTZ black hole.
On the other hand, using the Padmanabhan's approach, we can obtain the thermodynamic first law with the pressure \eqref{eq:P_BS} as $d \tilde{\cal E}_+= \tilde{{\cal T}}_+ d{{\cal S}_{\rm bs}^+}+\tilde{\Phi}_+ d{\cal Q}_{\rm tot} - {\cal P}d\tilde{{\cal V}}_+$ where
$\tilde{{\cal E}}_+ = 3{\cal M}_{\rm tot}/2 - {\cal M}_{\rm tot}^2/(2{\cal Q}_{\rm tot}^2)$.
However, the energy $\tilde{{\cal E}}_+$ is not the ADM mass, and  the temperature $\tilde{{\cal T}}_+$ and the potential $\tilde{\Phi}_+ $ are still different from the standard forms.

Now,
 we redefine the energy-momentum tensor  depending only on the cosmological constant
 by rearranging the equation of motion \eqref{eq:EOMpad1} and the energy-momentum tensor \eqref{eq:EOMpad2}
 as
\begin{align}
{\cal G}_{\mu\nu}&=R_{\mu\nu}-\frac{1}{2}g_{\mu\nu} R +2\nabla_\mu \nabla_\nu \phi -2g_{\mu\nu}\nabla^2 \phi +2g_{\mu\nu}(\nabla \phi)^2 +\frac{1}{24} g_{\mu\nu}H^2  -\frac{1}{4}H_{\mu\lambda\rho}H_{\nu}^{\lambda\rho}   \label{eq:calG}\\
&= \pi {\cal T}_{\mu\nu}, \notag  \\
{\cal T}_{\mu\nu} &=\frac{2}{\pi \ell^2} g_{\mu\nu}.\label{eq:calT}
\end{align}
From the energy-momentum tensor~\eqref{eq:calT},
we can simply read off the pressure as
\begin{equation}\label{eq:Pell_BS}
{\cal T}^{\hat{r}}_{\hat{r}}={\cal P}_\ell=\frac{2}{\pi \ell^2}.
\end{equation}
Note that
it will give desirable standard conjugate quantities,
for example, the ADM mass, the Hawking temperature, and so on,
which will be discussed in the next paragraph.

Using the equation of motion  ${\cal G}^{\hat{r}}_{\hat{r}}=\pi {\cal T}^{\hat{r}}_{\hat{r}}$  multiplied by $d\hat{r}_+$ in Eq.~\eqref{eq:calG} at the outer horizon,
we can obtain
\begin{equation}\label{eq:}
\frac{\pi^2 {\cal Q}_{\rm tot}^2+(2\hat{r}_+ -\hat{r}_-) \hat{r}_+^2 \hat{X}^2 f'(\hat{r}_+)}{\hat{r}_+^2 \hat{X}^2 \ell^2} d\hat{r}_+ = \pi {\cal T}^{\hat{r}}_{\hat{r}} d\hat{r}_+.
\end{equation}
 Then, the thermodynamic first law is derived as
\begin{align}
d{\cal M}_{\rm tot} &= \frac{\sqrt{{\cal M}_{\rm tot}^2-{\cal Q}_{\rm tot}^2}}{2\pi \ell {\cal M}_{\rm tot}} d{{\cal S}_{\rm bs}^+} + \frac{{\cal Q}_{\rm tot}}{{\cal M}_{\rm tot}} d{\cal Q}_{\rm tot}
+\frac{\pi \ell^2 ({\cal M}_{\rm tot}^2-{\cal Q}_{\rm tot}^2)}{4 {\cal M}_{\rm tot}}d{\cal P}_\ell    \nonumber \\
&= {\cal T}_+ d{{\cal S}_{\rm bs}^+}+\Phi_+ d{\cal Q}_{\rm tot} + {\cal V}_+d{\cal P}_\ell, \label{eq:Padell_BS}
\end{align}
where the potential $\Phi_+$ and the temperature ${\cal T}_+$ in Eq.~\eqref{eq:Padell_BS} are the same as
the potential \eqref{eq:Phi} of the axion charge from the two-form field
and the Hawking temperature \eqref{eq:TBS} from the surface gravity.
The conjugate variable ${\cal V}_+$ to the pressure can also be expressed by the comoving length of the black string as
${\cal V}_+=\pi^2 \ell \hat{X}^{-1}$.
As a result, it turns out that the pressure \eqref{eq:Pell_BS} is more plausible than the pressure \eqref{eq:P_BS}
in formulating the thermodynamic first law.

\subsection{Smarr relation from Christodoulou-Ruffini formula}

For asymptotically AdS black holes,
the cosmological constant is regarded as the bulk pressure in the black hole thermodynamics
in connection with the consistent formulation of the Smarr relation
in the framework of the scaling law from the Christodoulou-Ruffini formula,
while the thermodynamic first law is granted as universal \cite{Gibbons:2004ai}.
In this regard,
the thermodynamic first law and the Smarr relation should be formulated uniformly and consistently
by treating the cosmological constant as a thermodynamic variable \cite{Frassino:2015oca}.
Intriguingly,
the pressure-volume term appears
in spite of the asymptotic flatness in the black string
as seen from the thermodynamic first law \eqref{eq:Pad3_BS}.
Thus, we will examine how
the cosmological constant in the black string also leads to the pressure-volume term in the Smarr relation.

The Christodoulou-Ruffini formula of the black string is obtained
by inserting Eq.~\eqref{eq:Pell_BS} into  Eq.~\eqref{eq:entropy_X} as
\begin{equation}\label{eq:CRBS2}
{\cal M}_{\rm tot}=\sqrt{\frac{{{\cal S}^+_{\rm bs}}^2{\cal P}_\ell}{8\pi}+{\cal Q}_{\rm tot}^2},
\end{equation}
and it gives conjugate thermodynamic quantities $({\cal T}_+,\Phi_+,{\cal V}_+)$
to the thermodynamic quantities $({{\cal S}_{\rm bs}^+},{\cal Q}_{\rm tot},{\cal P}_\ell)$ as
\begin{align}
{\cal T}_+&=\left. \frac{\partial {\cal M}_{\rm tot}}{\partial {{\cal S}_{\rm bs}^+}}\right|_{{\cal Q}_{\rm tot}, {\cal P}_\ell}
=\frac{{{\cal S}_{\rm bs}^+}{\cal P}_\ell}{ \sqrt{8\pi( {{\cal S}_{\rm bs}^+}^2 {\cal P}_\ell+8\pi {\cal Q}_{\rm tot}^2)}} = \frac{\sqrt{{\cal M}_{\rm tot}^2-{\cal Q}_{\rm tot}^2}}{2\pi \ell {\cal M}_{\rm tot}}, \notag \\
\Phi_+ &=\left.\frac{\partial {\cal M}_{\rm tot}}{\partial {\cal Q}_{\rm tot}}\right|_{{{\cal S}_{\rm bs}^+},{\cal P}_\ell}
=\frac{\sqrt{8\pi} {\cal Q}_{\rm tot}}{\sqrt{{{\cal S}_{\rm bs}^+}^2{\cal P}_\ell+8\pi {\cal Q}_{\rm tot}^2}} =\frac{{\cal Q}_{\rm tot}}{{\cal M}_{\rm tot}}, \label{eq:intensive2_BS} \\
 {\cal V}_+ &=\left.\frac{\partial {\cal M}_{\rm tot}}{\partial {\cal P}_\ell}\right|_{{{\cal S}_{\rm bs}^+},{\cal Q}_{\rm tot}}=\frac{{{\cal S}_{\rm bs}^+}^2}{\sqrt{32\pi( {{\cal S}_{\rm bs}^+}^2 {\cal P}_\ell+8\pi {\cal Q}_{\rm tot}^2)}}
 =\frac{\pi \ell^2 ({\cal M}_{\rm tot}^2-{\cal Q}_{\rm tot}^2)}{4 {\cal M}_{\rm tot}}. \notag
 \end{align}
Then, we take the Christodoulou-Ruffini formula to be a homogeneous function of $\lambda^2 {\cal M}_{\rm tot}= {\cal M}_{\rm tot}(\lambda^3 {{\cal S}_{\rm bs}^+},\lambda^2 {\cal Q}_{\rm tot},\lambda^{-2} {\cal P}_\ell)$,
which results in the Smarr relation of the black string
as
\begin{equation}\label{eq:Smarr2_BS}
2{\cal M}_{\rm tot} = 3{\cal T}_+{\cal S}_{\rm bs}^+ +2\Phi_+{\cal Q}_{\rm tot}-2{\cal P}_\ell{\cal V}_+.
\end{equation}
The pressure-volume term appears in the Smarr relation even though the black string is asymptotically flat.
From the Smarr relation \eqref{eq:Smarr2_BS} with the definitions of the conjugate variables \eqref{eq:intensive2_BS},
it is straightforward to obtain the thermodynamic first law \eqref{eq:Padell_BS}.
Note that the Smarr relation corresponding to Eq.~\eqref{eq:Pad3_BS} can also be  read off as
$2{\cal M}_{\rm tot} = 3\tilde{{\cal T}}_+{\cal S}_{\rm bs}^+ +2\tilde{\Phi}_+{\cal Q}_{\rm tot}-2\tilde{{\cal V}}_+ {\cal P}$
from the Christodoulou-Ruffini formula of
${\cal M}_{\rm tot}=\sqrt{4\pi {\cal Q}_{\rm tot}^4/(4\pi {\cal Q}_{\rm tot}^2-{\cal P} {{\cal S}_{\rm bs}^+}^2)} $;
however, the conjugate quantities are different from the standard ones.

On the other hand, for the inner horizon,
the temperature from the surface gravity and the entropy from the Wald entropy \cite{Wald:1993nt}
vanish as
\begin{align}
{\cal T}_- =\left. \frac{\hat{r}_+}{2\pi \ell \hat{r}}\sqrt{1-\frac{\hat{r}_-}{\hat{r}}}~\right|_{\hat{r}=\hat{r}_-}=0, \qquad {\cal S}_{\rm bs}^-
=2\int \hat{r}_- \ell \sqrt{1-\frac{{\cal Q}^2}{{\cal M} \hat{r}_-}} d\hat{x} =0,\label{eq:entropy_BSin}
\end{align}
so that there is no thermal radiation and it seems to be meaningless to
discuss the thermodynamic first law and the Smarr relation at the inner horizon.

\section{Dual description of the thermodynamic quantities}
\label{sec:dual}

We find the dual relations of the thermodynamic quantities
between the BTZ black hole and the black string.
The dual invariance of the thermodynamic first law in terms of the quasi-local thermodynamic quantities per unit length was studied in Ref.~\cite{Ho:1997uk}.
However, in order to investigate the duality of the thermodynamic laws,
 it is convenient to employ the total mass and charge for the elementary black string,
 since the dual solution to the BTZ black hole is not the black string per unit length but the elementary black string.
Thus, we consider the dual relations of thermodynamic variables
of the BTZ black hole and the black string
 as $(M,S_+,J) \leftrightarrow ({\cal M}_{\rm tot},{{\cal S}_{\rm bs}^+},{\cal Q}_{\rm tot})$
and  $(T_+, \Omega_+,V_+) \leftrightarrow ({\cal T}_+, \Phi_+, {\cal V}_+)$
on the assumption of the relation of the pressures $P$ and ${\cal P}_\ell$.

The ADM mass ${\cal M}$ per unit length and
the axion charge ${\cal Q}$ per unit length of the black string are related to the ADM mass $M$ and
the angular momentum $J$ of the BTZ black hole as
${\cal M} = (1/2)\left(M\ell + \sqrt{M^2\ell^2-J^2}\right)$ and  $J^2=4{\cal Q}^2$ \cite{Horowitz:1993wt}.
If we assume the relation between the pressure of the BTZ black hole and that of the black string as $ P ={\cal P}_\ell/2$,
then we can find the dual relations of the thermodynamic variables between the BTZ black hole and the black string as
\begin{align}
M = \frac{{\cal M}_{\rm tot}^4-{\cal Q}_{\rm tot}^4}{4{\cal M}_{\rm tot}^2},  \quad
J = \frac{{\cal Q}_{\rm tot}\left({\cal M}_{\rm tot}^2 - {\cal Q}_{\rm tot}^2\right)}{\sqrt{2\pi {\cal P}_\ell }{\cal M}_{\rm tot}}, \quad S_+ = {\cal S}^+_{\rm bs}
 \label{eq:dual_th}
 \end{align}
with the dual relations of the conjugate variables as
\begin{align}
T_+ = \frac{{\cal M}_{\rm tot}^2-{\cal Q}_{\rm tot}^2}{2{\cal M}_{\rm tot}} {\cal T}_+,  \qquad
 \Omega_+ = \frac{1}{\ell}\Phi_+,  \qquad
 V_+ =  {\cal M}_{\rm tot}{\cal V}_+.  \label{eq:dual_th2}
\end{align}
Note that the timelike Killing vector was rescaled as $\partial/\partial t = \ell^{-1}(r_+^2-r_-^2)^{1/2} \partial/\partial \hat{t}$
 due to the rescaling of the time coordinate in order to make the dual metric to the BTZ black hole as the diagonalized form of the metric \eqref{eq:BS} \cite{Horowitz:1993wt}.
Thus, the Hawking temperatures are not the same but rather some multiple as
$T_+=\ell^{-1}(r_+^2-r_-^2)^{1/2} {\cal T}_+$.
Using the dual transformation \eqref{eq:dual_th} and \eqref{eq:dual_th2},
one can show that the thermodynamic relations of the BTZ black hole such as  the
thermodynamic first law \eqref{eq:first_BTZ}, the Christodoulou-Ruffini formula \eqref{eq:CRBTZ}, and the Smarr relation \eqref{eq:Smarr_BTZ} are transformed to those of the black string, Eqs.~\eqref{eq:Padell_BS}, \eqref{eq:CRBS2}, and \eqref{eq:Smarr2_BS}.

\section{Conclusion and Discussion}
\label{sec:sum}

In the BTZ black hole, we revisited the role of the cosmological constant
in the thermodynamic first law.
As expected, we confirmed that the negative temperature and the positive entropy appear
on the inner horizon of the BTZ black hole.
Our main result is
that, in the asymptotically flat black string, the cosmological constant also
gives rise to the pressure-volume term in the thermodynamic first law and the Smarr relation.
Consequently, it turned out that
the form of the pressure \eqref{eq:Pell_BS} written only in terms of the cosmological constant is preferable to the
complicated form of the pressure \eqref{eq:P_BS},
since the former pressure
gives the standard form of the conjugate variables in the thermodynamic first law and the Smarr relation
in contrast to the case of the latter pressure.
Finally, we showed that the thermodynamic quantities are well-behaved
under the dual transformation between the BTZ black hole and the black string.

Let us discuss  the energy-momentum tensor of the black string.
We identified the pressure \eqref{eq:P_BTZ} of the BTZ black hole by using the energy-momentum tensor \eqref{eq:EOMpad2} which consists of the dilaton field $\phi$, the three-form field strength $H_{\mu\nu\rho}$, and the cosmological constant $\Lambda$ as source terms.
However, the energy-momentum tensor \eqref{eq:EOMpad2} for the black string
gave the unusual pressure \eqref{eq:P_BS} which was incompatible with the standard definitions of the temperature and the potential,
whereas the pressure~\eqref{eq:Pell_BS}
identified from the energy-momentum tensor \eqref{eq:calT}
was compatible with the standard ones.
Therefore, this fact implies that
the dilaton field and the three-form field strength should be included in the gravitational part \eqref{eq:calG}
and the cosmological constant plays a role of the gravitational source~\eqref{eq:calT}.

The final comment is in order.
 There is another way to get the Smarr relation without the pressure
by considering the scale invariance of the reduced action for hairy black holes \cite{Banados:2005hm}.
Actually, authors in Ref.~\cite{Erices:2017nta} derived a Smarr relation for the BTZ black hole without the consideration of the cosmological constant as a thermodynamical variable as
$M=(1/2)T_+S_++\Omega_+ J$, which is different from the Smarr relation~\eqref{eq:Smarr_BTZ}
with the pressure.
However, one can obtain the Smarr relation without the pressure
from the Smarr relation \eqref{eq:Smarr_BTZ}
by rewriting the pressure-volume term into
the ADM mass and the temperature-entropy terms, {\it i.e.}, $2 P V_+=M+(1/2)T_+S_+$.
In this regard,
one can also convert the Smarr relation~\eqref{eq:Smarr2_BS} in the black string
into ${\cal M}_{\rm tot} = {\cal T}_+{\cal S}_{\rm bs}^+ +\Phi_+{\cal Q}_{\rm tot}$
without the pressure by
using the relation of $ {\cal P}_\ell {\cal V}_+=(1/2){\cal T}_+{\cal S}_{\rm bs}^+$.
Consequently, the Smarr relations from the Christodoulou-Ruffini formula are compatible with
the Smarr relations from the reduced action formalism.
\\
\\

\acknowledgments
We would like to thank Miguel Riquelme for very valuable comments on the manuscript, and thank Hwajin Um for exciting discussions.
This work was supported by
the National Research Foundation of Korea(NRF) grant funded by the
Korea government(MSIP) (2017R1A2B2006159).


\bibliographystyle{JHEP}       

\bibliography{references}

\end{document}